\documentclass[sigconf]{acmart}

\usepackage[frozencache]{minted}
\usepackage{xcolor}
\usepackage{amsmath}
\usepackage{graphicx}
\graphicspath{{figures/}}
\usepackage[pdf]{graphviz}
\usepackage{tikz}
\usepackage{pifont}
\usepackage{subcaption}
\usepackage{comment}
\usepackage{cleveref}
\usepackage{multicol}

\newcommand{\inlinehtml}[1]{\texttt{#1}}

\AtBeginDocument{%
  \providecommand\BibTeX{{%
    \normalfont B\kern-0.5em{\scshape i\kern-0.25em b}\kern-0.8em\TeX}}}

\copyrightyear{2024}
\acmYear{2024}
\setcopyright{acmlicensed}\acmConference[ITiCSE 2024]{Proceedings of the 2024 Innovation and Technology in Computer Science Education V. 1}{July 8--10, 2024}{Milan, Italy}
\acmBooktitle{Proceedings of the 2024 Innovation and Technology in Computer Science Education V. 1 (ITiCSE 2024), July 8--10, 2024, Milan, Italy}
\acmDOI{10.1145/3649217.3653599}
\acmISBN{979-8-4007-0600-4/24/07}

\usepackage{titlesec}
\titlespacing{\section}{2pt}{3pt plus 2pt minus 2pt}{3pt plus 2pt minus 2pt}

\newcommand{\tool}{FSM Builder}

\newcommand{\pl}{PrairieLearn}
\newcommand{\university}{University of Illinois Urbana-Champaign}

\begin{document}

\title{\tool: A Tool for Writing Autograded Finite Automata Questions}


\author{Eliot Wong Robson}
\orcid{0000-0002-1476-6715}
\affiliation{%
  \institution{\university{}}
  \country{Urbana, IL, USA}
}
\email{erobson2@illinois.edu}

\author{Sam Ruggerio}
\orcid{0009-0008-4163-8509}
\affiliation{%
  \institution{\university{}}
  \country{Urbana, IL, USA}
}
\email{samuelr6@illinois.edu}

\author{Jeff Erickson}
\orcid{0000-0002-5253-2282}
\affiliation{%
  \institution{\university{}}
  \country{Urbana, IL, USA}
}
\email{jeffe@illinois.edu}


\begin{abstract}
Deterministic and nondeterministic finite automata (DFAs and NFAs) are
abstract models of computation commonly taught in introductory computing theory
courses. These models have important applications (such as fast regular
expression matching), and are used to introduce formal language theory.
Undergraduate students often struggle with understanding
these models at first, due to the level of abstraction. As a result,
various pedagogical tools have been developed to allow students to practice with these models.

We introduce the FSM Builder, a new pedagogical tool enabling students to practice constructing DFAs and NFAs with a graphical editor, giving personalized feedback and partial credit. The algorithms used
for generating these are heavily inspired by previous works. The key advantages to
its competitors are greater flexibility and scalability. This is because the FSM Builder is implemented
using efficient algorithms from an open source package, allowing for easy extension and
question creation.

We discuss the implementation of the tool, how it stands out from previous tools, and takeaways from experiences of using the tool in multiple large courses. Survey results indicate the interface and feedback provided by the tool were useful to students.
\end{abstract}


\maketitle

\section{Introduction}
Formal languages and automata are a core topic in introductory computing theory courses
\cite{jtfcc-csc-13}. Undergraduate students often struggle with these concepts, as these courses
are usually the first exposure these students have
to abstract models of computation. The simplest of these models are Deterministic Finite Automata (DFA)
and Non-Deterministic Finite Automata (NFA), which are used to define regular languages. DFAs and NFAs are
classified as \emph{finite-state machines} (FSMs), or (equivalently) finite automata. 

Many assignments require students to design FSMs to match English descriptions of regular languages or regular expressions. In our experience, students have frequently asked for more of these types of practice problems, along with feedback to their solutions. In a large introductory theory course, providing meaningful one-on-one feedback for a multitude of practice problems is infeasible, and the time needed to provide this feedback prevents many students from engaging fully with these topics. Thus, providing practice with automated feedback represents
a significant opportunity to increase student engagement while also easing the workload of course staff.

To meet these demands, we developed the \tool{} as a successor to previous finite automata tools, implementing similar key features while also providing functionality to aid in development of new exercises at scale for large courses. This tool was
developed for easy integration with the \pl{} platform \cite{prairielearn}, but 
the implementation is self-contained and may be used elsewhere.

\subsection{Finite Automata Background}
We briefly review standard terminology for FSMs \cite{Sipser12}.
A finite state machine consists of an input alphabet (the characters that can be
used in input strings), a set of states, a set of accepting states, a start state, and a transition function. The transition function defines a state to transition to from every state and every character from the alphabet (unless explicitly stated that missing characters go
to an implicit dump state, which by convention results in rejection).
The set of accepted strings is the language accepted by the FSM, and there may
be multiple different FSMs that accept the same language.

Of note is that DFAs cannot have multiple transitions leaving a given state on the same character, while NFAs can (as this is how nondeterminism is incorporated). Additionally,
NFAs may contain epsilon transitions.

\subsection{Organization}
The rest of the paper is organized as follows. We first review the features and limitations of other tools in \Cref{sec:related_work}. We outline design considerations for the \tool{} in \Cref{sec:design_considerations} and provide background for the course and development of the tool in
\Cref{sec:course_context}. In \Cref{sec:user_interface,sec:autograder}, we discuss the features
of the user interface and backend autograder respectively, including the specific algorithms used to 
generate partial credit and feedback. In \Cref{sec:evaluation} we present survey results from using
the \tool{} in an undergraduate algorithms course with over 300 students. \Cref{sec:adoption} provides
details on how to adopt the tool, and \Cref{sec:limitations_and_future_work} concludes with a discussion of future directions.

\section{Related Work}
\label{sec:related_work}
There has been substantial prior work on software tools enabling automated teaching of automata
theory, with a specific focus on finite automata \cite{csk-fyasr-11}. In this section, we discuss the
strengths and weaknesses of previous tools and compare to those of the \tool{}.

\subsection{JFLAP}
One of the earliest such tools is the
JFLAP software package \cite{Rodger06, rwlmos-ieatj-09}, which allows interactive exploration of automata
by students, including the simulation of input strings. The original aim of JFLAP was to give students
an interactive way to explore course content, as the original package does not provide mechanisms for feedback or evaluation by instructors. Moreover, the original package required a local installation, making
scaling to a large course difficult. JFLAP has support for more types of automata than just finite
automata \cite{cfr-viatcj-04}, which is beyond the scope of the \tool{}.

\subsubsection{DAVID Extension}
A follow-up work on JFLAP analyzed the impact of automated
feedback on the student experience. Specifically, \citet{bfhmn-esf-22}
developed the DAVID extension, which provides automated feedback through counterexamples, and analyzed
the effect on student performance. This feedback, called a \textit{witness string}, is the
shortest string where the student submission and reference solution have differing behavior. The feedback mechanism in the \tool{} is an extension of this idea, providing multiple such strings if possible.

Although the DAVID extension used an autograder in addition to the feedback provided, this
autograder did not provide partial credit.

\subsubsection{OpenFLAP}
More recently, the functionality in JFLAP was expanded with an autograder in OpenFLAP \cite{msr-tflvage-21}
as part of an effort to develop an eTextbook for automata theory. This tool can be integrated with
existing learning management systems, but cannot be integrated with \pl{} easily. In addition, the
autograding algorithm is not as robust as that of the DAVID extension, using test strings instead
of analyzing the FSMs for equivalence.

\subsection{Automata Tutor}

More recently, Automata Tutor has emerged as a similar tool to JFLAP, providing a web-based interface
with a greater emphasis on graded assessments \cite{DAntoni15}. In particular, the most recent
version of Automata Tutor provides automated feedback for a number of different question types related
to automata (not just FSMs), with a focus on large courses \cite{DAntoni20}. However,
this tool cannot be integrated with \pl{} easily, and does not seem to allow for custom grader code the way the \tool{} does.

\subsection{FSM Designer}
\label{sec:fsm_designer}
The FSM Designer \cite{Wallace15} is a graphical user interface for typesetting automata and does not feature any kind of simulation or autograding.
Although the FSM designer has a narrower scope than other tools, it has an intuitive interface, freely available source code, and was already a popular tool many students were familiar with. The user interface of the \tool{} is based on that of the FSM Designer to facilitate easy adoption by students, and encourage students to typeset their homework by giving them familiarity with the FSM Designer.

\subsection{Partial Credit}
\citet{AlurDGKV13} analyzed different partial credit schemes for DFA construction questions.
Although not directly compared to other partial credit schemes, the density difference partial credit scheme
proved to be the easiest to integrate into the \tool{}.

\section{Design Considerations}
\label{sec:design_considerations}
Some desirable features of the previously discussed tools incorporated into the \tool{} are the following:

\begin{enumerate}
    \item A simple, modern graphical user interface for students. See \Cref{sec:student_interface}.

    \item A robust grading algorithm that can efficiently check whether a student submission is equivalent to a reference solution and give partial credit. See \Cref{sec:check_correctness,sec:partial_credit}.

    \item A string-based feedback mechanism that can generate counterexamples to student submissions. See \Cref{sec:feedback}.

    \item The ability to create questions and practice assignments. 
\end{enumerate}

To distinguish itself from other tools, the \tool{} incorporates
the following unique features:

\begin{enumerate}
    \item Compatibility with \pl{}, the course content hosting platform.
    \item Reinforcement of course conventions for designing more human-readable FSMs. See \Cref{sec:conventions}.
    \item Fast creation of practice problems from existing reference solutions with
    minimal developer overhead (no need for custom grader code if using the included autograder). See \Cref{sec:instructor_interface}.
    \item Use of the graphical interface independently of the autograder (for questions with custom grading algorithms).
    \item A self-contained, modular, open source implementation in Python and JavaScript. Importantly, all code from the \tool{} can be integrated into other content hosting platforms if desired. 
\end{enumerate}

The above lend to the scalability and flexibility of the \tool{}, 
which are the key characteristics distinguishing it from prior work.

\section{Course Context}
\label{sec:course_context}
The \tool{} was developed and pilot tested as part of the introductory algorithms and models of computation course in the computer science department
at the \university{}. Most students are second year computer science majors taking the course as a degree requirement. This is a large course, with enrollment between 370 and 400 students each semester. The models of computation portion
of the course spends a considerable amount of time on finite state machines and regular languages. The first author of this paper has been the lead developer for the course
since Fall 2021, and the second author has been a developer since Spring 2022. The
third author has been a lead instructor for the course and written much of the
course material.

One of the most common questions asked of students on finite automata is, given an English description of a language, prove this language is regular by providing a finite state machine accepting this language. This question structure is commonly used in homework
and exam questions throughout the course, and accordingly, most of the existing practice questions for regular languages follow this structure. We designed the interface for question
writing to facilitate efficient creation from existing solutions to
these types of problems.

The course also uses the \pl{} platform \cite{whz-pl-15} to deliver other types of course content, so any tool used needed to be compatible with this platform. See \cite{pl4tcs} for more details on content development for our course.

\subsection{Tool Development}
Prior to the development of the \tool{}, our course used a system for autograding FSMs that required programmatic input from both students and question writers, hosted on \pl{}. Students were required to write a short Python program defining a finite automata, and
this was checked against the reference solution using an
equality checking algorithm based on the product construction for DFAs \cite{Sipser12}. 
During these semesters, a common sentiment was the desire for a visual editor that did not require writing Python code.

The key motivation for the creation of the \tool{}, rather than using existing
tools, was desire for integration with \pl{}.
With the design considerations outlined in \Cref{sec:design_considerations}, during the summer of 2022, we wrote the first proper version
of the \tool{}. This included the full graphical interface based on the FSM designer, a simple 
interface for question writers,
and string based feedback.

In summer of 2023, we updated the \tool{} with another round of feedback from course
instructors, incorporating the language-similarity based partial credit scheme, and more detailed feedback for incorrect answers by students. With these new features in place, the tool has seen adoption in other courses, including 
the introductory discrete mathematics course within the same department.

\section{User Interface}
\label{sec:user_interface}
The \tool{} is integrated within \pl{}, providing an intuitive user experience for students and an interface for question writers that requires minimal knowledge of automata theory. Of particular importance is the ease of question creation using the included autograder, as this facilitated the production of a large number of exercises in a short period of time from existing reference solutions.

\subsection{Student Interface}
\label{sec:student_interface}

\begin{figure*}
\begin{subfigure}[b]{0.55\textwidth}
\includegraphics[width=\textwidth]{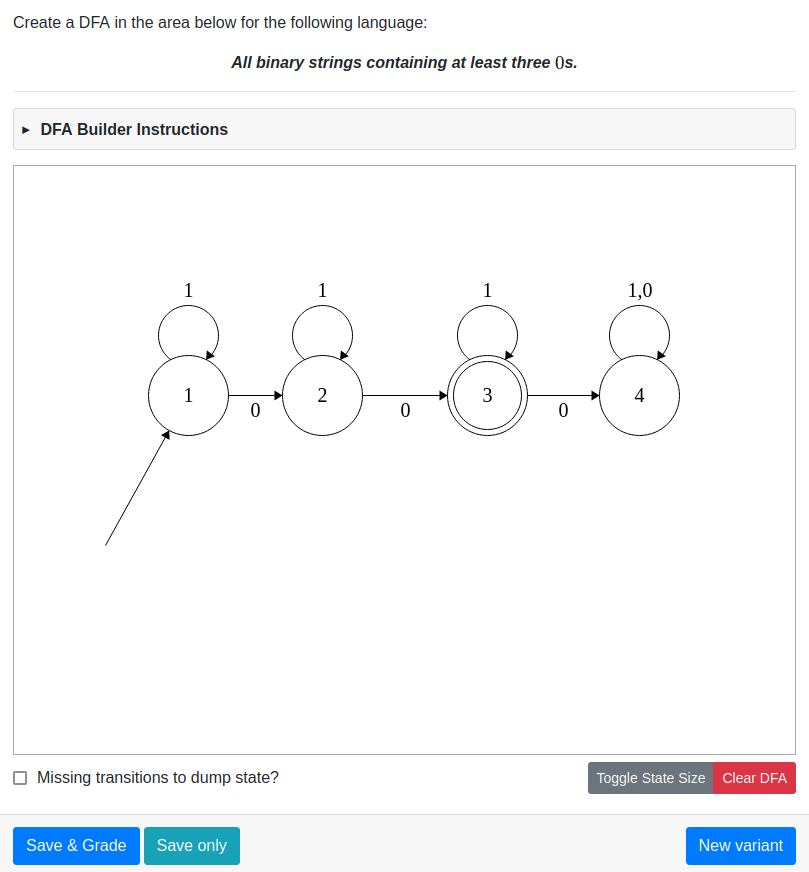}
\caption{The student interface for the \tool{}. Note the student submission in the figure does not match the desired language.}
\label{fig:canvas}
\end{subfigure}
\hfill
\begin{subfigure}[b]{0.4\textwidth}
\includegraphics[width=\textwidth]{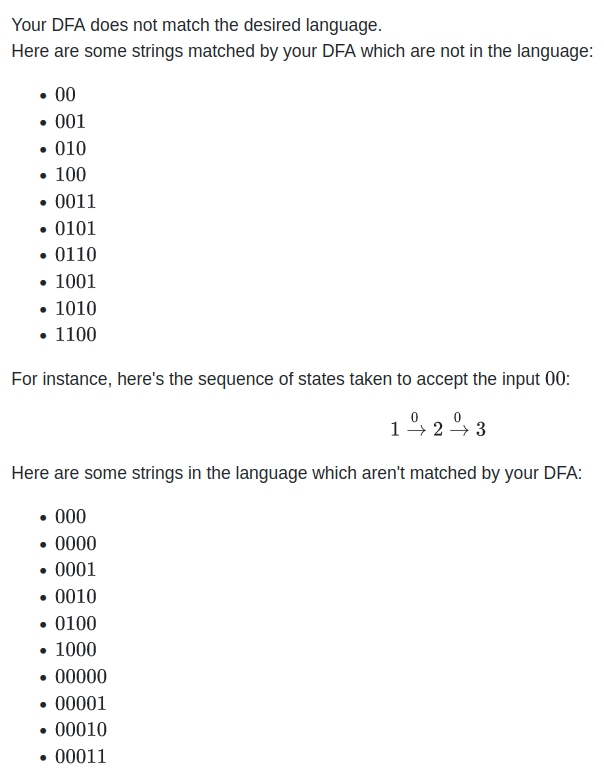}
\caption{Feedback shown to the student for the submission on the left.}
\label{fig:feedback}
\end{subfigure}
\caption{An incorrect student submission in the \tool{} and corresponding feedback.}
\label{fig:fsm}
\end{figure*}

As alluded to in \Cref{sec:fsm_designer}, the interface for students is
based on the FSM Designer by \citet{Wallace15}.
Students are given a canvas where they can add new states and transitions to their finite state machine by clicking. States can be moved once created and transitions between states can be repositioned to keep the drawing clear. States can be given labels consisting of any characters, and transitions can be labeled with characters from the language's alphabet. An instance of the \tool{} is shown in \Cref{fig:canvas}.

Students receive automated feedback from their response both on the canvas and feedback box below once they select "Save \& Grade".

\subsection{Instructor Interface}
\label{sec:instructor_interface}
In \pl{}, an instructor can create a new question using the custom HTML element for the \tool{}. This HTML element takes a JSON object defining a DFA or NFA used as the correct answer. This JSON object contains the states, alphabet, transitions, initial state, and accepting states. The following is an example JSON object used for the reference solution in \Cref{fig:fsm}:
\begin{minted}{json}
{
  "states": ["0", "1", "2", "3"],
  "input_symbols": ["0", "1"],
  "transitions":{
      "0":{"0": "1", "1": "0"},
      "1":{"0": "2", "1": "1"},
      "2":{"0": "3", "1": "2"},
      "3":{"0": "3", "1": "3"}
  },
  "initial_state": "0",
  "final_states": ["3"]
}
\end{minted}

Instructors can either write JSON directly when creating new questions, or use the \tool{} itself to generate the JSON for the reference FSM. Unlike the previous solution we had for building questions with Python, there is no scripting necessary
to use the autograder, only static HTML elements. For general information
on custom elements, see \cite{plelementdocs}.

\section{Autograder}
\label{sec:autograder}
The \tool{} uses autograding features that both reinforce clear writing conventions for designing FSMs, and check for correctness against a reference solution using finite automata based algorithms. Note that the autograder is tailored to the most common question format used by our course, asking students to provide an FSM that accepts a given target language.

\subsection{Enforcing FSM conventions}
\label{sec:conventions}
Before any in-depth feedback on the language of an input automaton is given by the grading algorithm,
the student submissions are checked that they define a valid automaton (DFA or NFA)
according to the following conventions:

\begin{enumerate}
    \item All provided state names are nonempty and unique.
    \item Exactly one start state is marked.
    \item The FSM does not have any non-accessible states.
    \item All transitions are defined on valid characters in the alphabet.
    \item The automaton must have transitions leaving every state for every character, unless specified that missing transitions go to a dump state.
    \item For a DFA, the submission should not have
    multiple transitions leaving a given state on the same character.
\end{enumerate}

Using these checks ensures that the FSM is well defined, and helps promote
clear writing. Where possible, parts of the submissions that must
be corrected according to the above rules are highlighted in red on the canvas. This gives personalized feedback on specific errors in a submission, similar to what students might get from course staff.

\subsection{Checking correctness of an automaton}
\label{sec:check_correctness}
Once the student submission has been validated based on the above criteria, it is converted by the grader code into a Python object representing the automaton. The grader code creates a similar object for the automaton provided as a reference solution, then compares them for language equality to award full credit. Note that this type of comparison is robust to different underlying FSMs, meaning that any student submission accepting the same language as the reference solution will be given full credit by the autograder.

This comparison is done with an optimized version of the Hopcroft-Karp algorithm \cite{Almeida2009TestingTE, Norton09}. Importantly, this algorithm has nearly-linear runtime in the size of the input, and is easy to implement using standard data structures, making it very well-suited for this application. The runtime was of particular importance, as the size of student submissions necessitated the use of an algorithm more efficient than the standard product construction algorithm \cite{Sipser12}. This is the same comparison algorithm used by \citet{bfhmn-esf-22}.

\subsection{Partial Credit}
\label{sec:partial_credit}

If the student submission was not awarded full credit by the equivalence algorithm, we use the language-based partial credit scheme described by \citet[Section 3.3]{AlurDGKV13}, called the \emph{approximated density difference}. In detail, for two regular languages $L_1, L_2 \subset \Sigma^*$, this quantity is defined as
\[
    \textsc{A-Den-Dif}(L_1, L_2)
    \coloneqq%
    \frac{1}{2k+1} \sum_{n=0}^{2k} \frac{\left| (L_1 \oplus L_2) \cap \Sigma^{n} \right|}{\max(\left| L_2 \cap \Sigma^n \right|, 1)}
\]
where $k$ is the number of states in the minimal DFA representing $L_2$. In our grading algorithm, $L_2$ is the language of the reference solution. The expression contained in the sum is the number of strings of length $n$ misclassified by $L_1$ scaled by the number of strings of length $n$ accepted by $L_2$. Intuitively, this is the size of the discrepancy between $L_1$ and $L_2$ for a given length $n$, which is then summed over all lengths up to $2k$.

This partial credit algorithm was chosen because it was practical to implement efficiently using the primitives provided by the automata package \cite{Evans_automata_A_Python_2023}, and conclusions by \citet{AlurDGKV13} stating that this algorithm performed well in cases where a student answer misclassified a small, finite number of strings (for example, only misclassifying the empty string). 

\subsection{Feedback}
\label{sec:feedback}
In addition to the partial credit scheme, the grading algorithm also generates feedback from strings misclassified by the student's FSM. All strings
up to length 8 are checked whether they were incorrectly accepted or incorrectly rejected
by the student's FSM, and then the (lexicographically) first 10 are given as feedback to the student. If no misclassified strings are found during this search, then a minimal length misclassified string is generated using an extension of the DFA equality algorithm \cite{Norton09}. For the first string incorrectly accepted by the student's FSM, the sequence of states taken to reach an accepting state is shown. An example of this feedback is shown in \Cref{fig:feedback}.

This can be viewed as an expanded version of the feedback system used by the DAVID extension \cite{bfhmn-esf-22}, where instead of a single string witnessing that the student's submission is not equal to the reference solution, we give multiple such witness strings as feedback if possible. This expanded feedback allows students to identify patterns in misclassified strings that can more quickly lead to finding a correct solution.

\subsection{Implementation}

The autograder is written in Python and uses the automata package \cite{Evans_automata_A_Python_2023} to implement to the main grading and
feedback algorithms. The package provides robust primitives for efficient manipulation of regular languages, making it easy to write custom grader code for other types of assessments involving finite automata. This means the \tool{}
is very self-contained (the automata package is the only major dependency) and more extensible than previous tools.

In particular, the package provides optimized algorithms for converting regular expressions to NFAs, NFAs to DFAs, minimizing DFAs, enumerating the strings belonging to the language of a DFA, and the product construction for DFAs. These subroutines were critical in writing the grading and feedback algorithms for the \tool{}, and represent the most technically challenging parts of the implementation. The availability of these subroutines opens the door for the creation of more custom questions using the tool.

\section{Evaluation}
\label{sec:evaluation}
\subsection{Scalability}
A key motivating factor in the development of \tool{} was scalability to both
large courses and large numbers of questions. The \tool{} has been very successful on
these fronts, as in our course, we were able to convert our entire backlog of finite automata questions (over 50) to automated practice problems without the need to write custom
grader code for any individual question (only using the autograding functionality
described in \Cref{sec:autograder}). The tool scaled well to the questions
themselves, as we did not encounter any efficiency issues with the grading or feedback
algorithms, and the tool integrated cleanly with \pl{}. This scalability has been observed
in other courses, as the large introductory discrete mathematics course in our department has also begun
using the \tool{}.

The \tool{} has also proven to be extensible and flexible, as we were able to develop multiple questions using custom grading algorithms.

\subsection{Student Response}
To evaluate student responses to the \tool{}, we conducted a voluntary, fully anonymous online
survey distributed to students in the introductory computing theory course during the 
Fall 2023 semester. The survey consisted of Likert scale questions about the experience of using the tool. All students had interacted with the tool as part of a short (required) homework question. Of the 383 enrolled students, 246 students had additionally used the tool as part of optional review content, and 196 responded to the survey. Students were incentivized by granting the
entire course a small amount of extra credit if over half of students in the course completed the survey. The survey platform prevented students from submitting responses more than once.

The \university{} IRB office gives this survey a non-human subjects research designation, as all of the data collection was completely anonymous. 

The goal of the survey was to assess general impressions of the user interface and feedback of the tool in comparison to feedback given by course staff. The questions were developed
to assess the general sentiment towards different aspects of the tool, rather than doing
detailed comparisons with other tools. Importantly, the \tool{} was not being used to
replace any existing course content, so our evaluation focused on whether students
found the content delivered through the tool useful and engaging.

\begin{figure*}
\centering
\includegraphics[width=\textwidth]{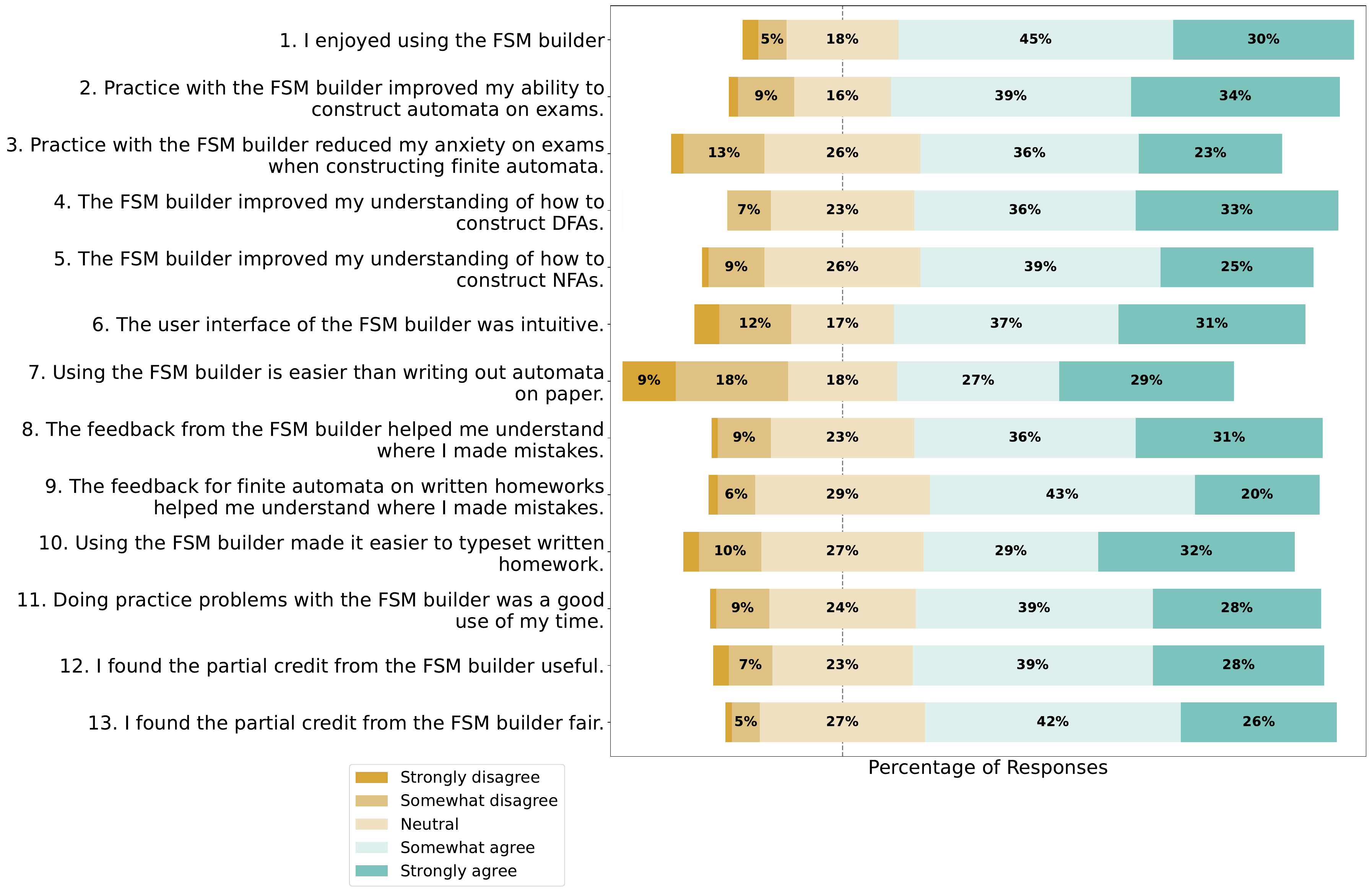}
\caption{Responses to the Likert scale questions on the survey. There were 196 total responses.
}
\label{fig:survey_results}
\end{figure*}

The results of the survey are shown in \Cref{fig:survey_results}, and overall feedback to the \tool{} was very positive. Notably, there were more students who agreed the \tool{}
provided useful feedback (statement 8, 131 positive responses) than students who agreed written homework graded by a TA provided useful feedback (statement 9, 125 positive responses). While not conclusive evidence feedback by the tool was superior, this demonstrates students were generally satisfied with the automated feedback, indicating the \tool{} is able to provide valuable feedback to students at scale.

The statement most disagreed with was number 9, that it is easier to use the \tool{} than to construct an automaton on paper. However, more respondents agreed than disagreed, and this mixed response indicates that using the tool is still advantageous for large courses, as grading automata on paper does not scale well among many students. Despite this, 67\% of respondents (131 out of 196) agreed the user interface of the \tool{} was intuitive (statement 6).

Students also responded positively to statement 10 (61\% agreed, 119 out of 196), that the tool made it easier to typeset written homework. This provides some positive evidence for
the choice to use the FSM Designer as the basis for the user interface.

\section{Adopting the \tool{}}
\label{sec:adoption}
To try the FSM builder, look at the DFA and NFA practice assessments
available in our public course instance \footnote{\url{https://www.prairielearn.org/pl/course_instance/129595}}.

To use the \tool{} in your course, start by following the onboarding
instructions for \pl{} \cite{pldocs}. Next, follow the directions in this repository \footnote{\url{https://github.com/eliotwrobson/FSMBuilder}}
on integrating the FSM builder with your course. Note that the code for this implementation is self-contained.

\section{Limitations and Future Work}
\label{sec:limitations_and_future_work}
The key limitation of this work is the scope for the evaluation of the \tool{}.
Most of the questions students practiced on were part of optional review content and
of very similar format, and completion of the survey was voluntary. The results
from the survey were positive, but the nature of the evaluation means we can only
draw general conclusions. Although we found this acceptable for an initial investigation, more detailed data is required to draw stronger conclusions on the effectiveness of the \tool{}.

With a variety of different independent components, the \tool{} provides a number of avenues for
future research. Specifically, the graphical interface shown to students and the backend
grader code are modular components, and further work could explore the student response to
different types of partial credit, different feedback schemes,
and potentially a different student interface to that of \citet{Wallace15}.

Furthermore, the backend grader code used by the \tool{} can be easily adapted to work with regular expressions, using subroutines in the automata package \cite{Evans_automata_A_Python_2023}. We have such a tool in our course, providing nearly identical feedback to that of the \tool{}. Possible future work could examine the effectiveness of this regular expression tool.

Orthogonal to altering the \tool{} itself, most of the problems completed by students were in very similar assessment contexts. A direction for further work
could examine student responses to the partial credit and feedback schemes when used for different types
of questions (including questions that involve randomization and automated generation of question prompts), and in different assessment contexts, such as exams.

\begin{acks}
We thank Seth Poulsen and Yael Gerter for suggestions and feedback on the survey questions we used. We would also like to thank Morgan Fong, Yael Gerter, Seth Poulsen, and the Computers \& Education research group at the \university{} for feedback on earlier versions of this paper

Funding for this work was partially provided by the Strategic Instructional Innovations Program at the \university{}.
\end{acks}

\onecolumn

\begin{multicols}{2}
\bibliographystyle{ACM-Reference-Format}
\bibliography{paper-fsm-builder-tool}
\end{multicols}
\end{document}